# Computer simulations of the interface zone structure in binary eutectic

O. Bystrenko, V. Kartuzov

Frantsevich Institute for material science problems, Kiev, Ukraine
e-mail: obystr@bitp.kiev.ua

**Abstract**

Molecular dynamics simulations of the interface structure in binary AgCu eutectic were performed by using the realistic EAM potential. In simulations, we examined such quantities as the time dependence of the total energy in the process of equilibration, the probability distributions, the composition profiles for the components, and the component diffusivities within the interface zone. It is shown that the equilibrium in the solid state associated with the complete phase separation in binary eutectic is accompanied by formation of the steady disordered diffusion zone at the boundary of the crystalline components. At higher temperatures, closer to the eutectic point, the increase in the width of the steady diffusion zone is observed. The particle diffusivities grow therewith to the numbers typical for the liquid phase. Above the eutectic point, the steady zone does not form, instead, the complete contact melting in the system occurs. The results of simulations indicate that during the temperature increase the phenomenon of contact melting is preceded by the similar process spatially localized in the vicinity of the interface.

**Introduction**

The properties of eutectic systems are of interest since they are abundant in nature and widely used in industrial processes such as sintering in powder metallurgy, synthesis of composite materials for electronics, photonics, etc. (see, for instance, [1-4]). One of central issues essential for understanding of the processes of structure formation in eutectic systems are the properties of interfaces between the phases, in particular, their behavior near the eutectic temperature including the phenomenon of contact melting (CM).

Binary eutectic systems have been much studied, and, in general, their behavior is well understood [5-7]. According to the commonly accepted opinion, the thermal equilibrium in binary eutectic in solid state is associated with the complete decomposition into the ideal crystal phases with coherent or semi-coherent interface of zero width. The eutectic point on the phase diagram is associated therewith with the equilibrium of three phases. However, the experiments with the formation of diffusion zone and the phenomenon of contact melting in binary eutectic [8-9], as well as the simulations of the properties of crystal interfaces for various systems (see, for instance, the review [10] and the references therein) indicate that, in actual fact, the equilibrium interface between the crystal phases may have the non-zero width. In this case the equilibrium interface width and energy for binary eutectic acquire the meaning of physical quantities which have to be determined from the experiment. Significant argument in favor of this suggestion is the reduce in the width of the diffusion zone observed with lowering the temperature in the above experiments as well as known enhanced stability of eutectic interfaces with regard to the diffusion. This fact is used, for instance, in manufacturing X-ray mirrors or composite materials.

The theoretical interpretation for the experimental results of Refs. [8-9] based on the standard model of the phase-field theory (PFT) was suggested in the previous work of the authors [11], where the contact phenomena in binary eutectic are explained by the effect of the interface energy and its contribution to the total free energy of the system. It is to mention, that this mechanism predicts the formation of equilibrium three-phase (solid-liquid-solid) states in eutectic systems in the vicinity of the eutectic point. Such states can be regarded, on the one hand, as the states of complete decomposition with the liquid interface of finite width, and, on the other hand, they suggest the explanation of the phenomenon of phase separation observed in liquid eutectic above the eutectic temperature [6]. The triple points gets therewith "blurred", and, as the consequence, each of solid phases is in equilibrium separately with the liquid interface layer.

The above issues are of importance since they concern the commonly accepted concept of the equilibrium of solid crystal phases and structure formation in eutectic. For instance, one of the arguments in favor of the necessity to employ the multi-phase PFT model for binary eutectic is exactly the fact that the standard PFT model fails to describe the zero-width interface with finite energy as well as the properties of the triple point [12].

It is clear, that the PFT model used in Ref. [11] represents, in fact, a phenomenological approach which takes into account only the basic thermodynamic properties of the system, while the microscopic component structure and the associated critical and fluctuation phenomena remain ignored. For this reason, it seems important to study the above issues on the basis of microscopic atomistic treatments.

In this work we examine theoretically the behavior of the interface between the phases, by taking as an example the binary AgCu eutectic. The study is based on molecular dynamics (MD) simulations with the use of the EAM-potential proposed in Ref. [13]. The AgCu system is experimentally well examined and distinguished by the simple phase diagram [14], while the above mentioned potential reproduces the latter in wide range of compositions and temperatures. Therefore, it can provide the basis for realistic numerical experiments. The goal of simulations was to study the relaxation kinetics of the interface between the ideal crystals to the equilibrium, examine its structure near the eutectic temperature, and investigate the intensity of the diffusion of the atoms within the boundary interface zone.

**Numerical simulations. Results and discussion**

Computer simulations of contact phenomena in AgCu system were performed by means of MD method with the use of the free program package LAMMPS [15]. The employed effective EAM-potential reproduces qualitatively phase diagram for the system as a whole. The theoretical numbers for eutectic temperature and concentration given by the authors of Ref. [13] are $T_{eut} = 935°K$ and $c_{eut} = 0.46\ (at.Ag)$, whereas the respective experimental values are $T_{eut}^{exp} = 1052°K$, $c_{eut}^{exp} = 0.6\ (at.Ag)$[14].

The parameters used in simulations were specified as follows (below we use angstrom as the length unit; the mass is measured in atomic units): the lattice type FCC (for both components); lattice parameters $a_{Cu} = 3.61$, $a_{Ag} = 4.085$; atomic masses $m_{Cu} = 63.5$, $m_{Ag} = 107.9$.

The initial system configuration was set as two adjacent rectangular parallelepipeds filled with crystal lattices of copper and silver. The silver lattice was rotated in XOY plane in relation to copper lattice at the angle $\phi = 45°$, provided that the OX axis direction is specified to be perpendicular to the interface plane (Fig.1). The initial separation between the outside atomic planes of copper and silver, which form the interface, was set in such a way as to minimize the initial perturbation in the system, and was determined as $d=1.44$. It is to mention, that we tried several possible mutual orientations of lattices. The above specified orientation was chosen to be the most relevant, since it provided the minimum temperature of contact melting (close to the theoretical eutectic point).

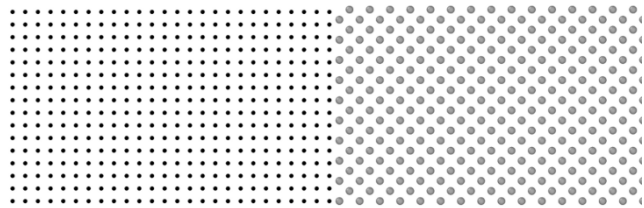

**Fig.1** Initial configuration. The positions of Cu and Ag atoms are specified as the small black and large grey dots, respectively

The simulations were performed for NPT-ensemble with the use of periodic boundary conditions, for the pressure P=1 bar. The total particle number in the system was $N=8688$ ($N_{Cu} = 5328$; $N_{Ag} = 3360$.

We carried out a series of numerical experiments for different temperatures in the range $T = 600 \div 1050°K$. In Fig.2, the final particle configurations obtained in simulations and the energy behavior as a function of time in the process of the relaxation in the system for the temperatures $T = 850; 900; 950; 1000°K$ are displayed. As is seen from the results, for the temperatures $T = 850; 900°K$, below the theoretical eutectic point $T_{eut} = 935°K$, the energy gets stabilized during relatively short time interval $t \simeq 50ns$, and the system comes to a steady state.

The associated particle configurations demonstrate that the interaction between the crystal components gives rise to the formation of the disordered interface (diffusion) zone with the width of 5-20 angstrom.

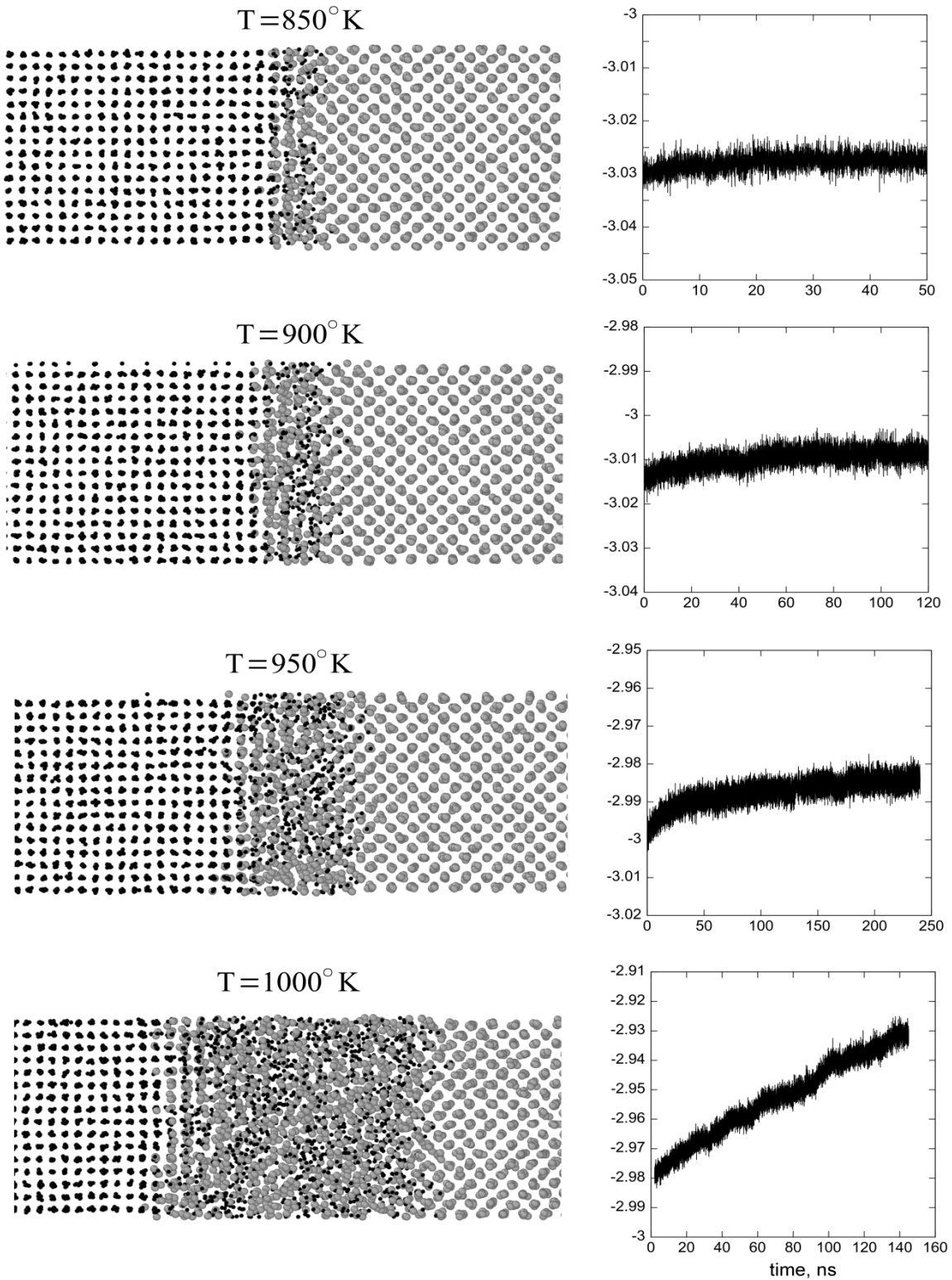

**Fig.2** Particle configurations in the AgCu system obtained in simulations (left) and the dependence of the total energy per particle (eV) on time in the process of the relaxation (right) for the temperatures $T = 850; 900; 950; 1000°K$ (from the top to the bottom). Small black and large grey dots stand for Cu and Ag atoms, respectively

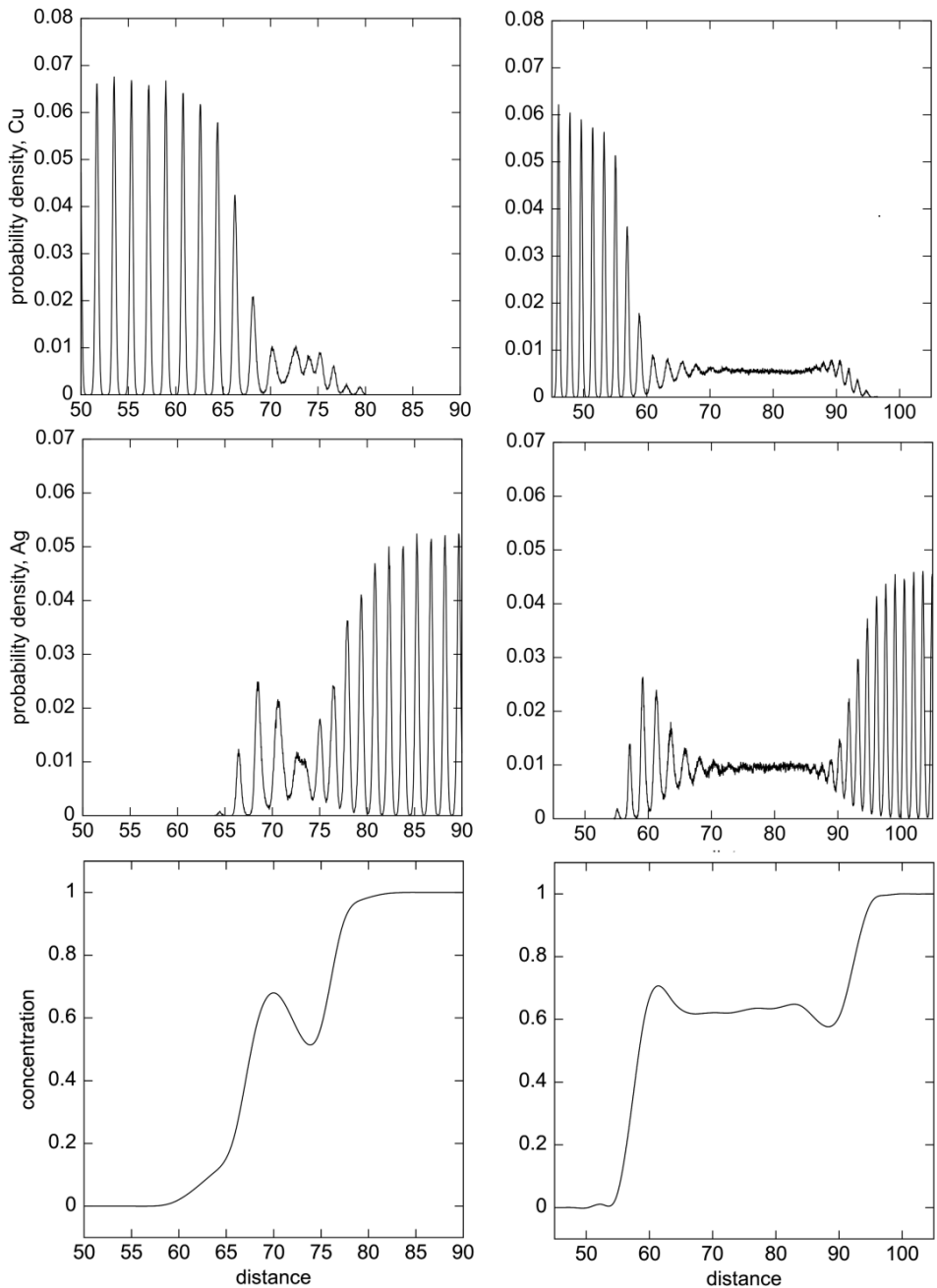

**Fig.3** Probability distributions for Cu and Ag atoms (top and middle) and concentration profiles (bottom) obtained in simulations near the interface zone for the temperatures $T = 900°K$ (left) and $T = 1000°K$ (right)

The observed processes show that the initial configuration of the adjacent ideal crystal lattices is non-equilibrium. The free energy of the system can only decrease in the process of relaxation to the equilibrium, therefore, some increase in the total energy indicates that the entropy contribution due to the loss of order in the diffusion zone is dominating. It is worth mentioning that, in any event, the thermal equilibrium in solid binary eutectic is associated with the complete decomposition into the components. Therefore, it is natural to assume that the steady states obtained in simulations are the ones describing the equilibrium between the crystal silver and copper phases (at least for the given component orientation). It is to note, that lowering the temperature results in reduction of the contact phenomena (such as the extent of disorder, the change in energy in the process of relaxation, and the width of the diffusion zone), and for the temperatures much lower than the eutectic one, they may be insignificant.

For the temperatures above the eutectic point ($T = 950; 1000°K$), the total energy does not stabilize, instead, the regime of the linear growth with respect to time sets in. The width of the diffusion zone continues to grow as well. It is clear that in this case there occurs the phenomenon of CM in the system, and in the end one can observe the complete melting of components.

The calculated probability distributions (unary distribution functions) for the copper and silver atoms and the concentration profiles (smoothed by the averaging over two lattice periods) in the vicinity of the interface are given in Fig.3. As is seen from it, for the temperature below the eutectic point, the amplitude of atomic oscillations within the interface zone increases (which follows from the broadening of the probability peaks associated with the atomic planes), and the significant increase of disorder occurs. At the same time, the correlations in the atomic positions in OX direction (perpendicular to the interface) still remain pronounced. Above the eutectic temperature, on the beginning of the CM process, the correlations along OX axis are absent, which suggests the formation of the liquid layer. The non-monotonical shape of the concentration profiles should be pointed out, which can be viewed as the manifestation of anisotropic phase separation in eutectic in OX direction.

Transport properties is another important feature of the processes occurring in the interface zone. In order to estimate the intensity of the diffusion of copper atoms in the boundary layer, we performed a series of simulations in the following way. The three-dimensional diffusion coefficient has been determined from the Einstein's relation

$$D = \frac{\langle \Delta r^2 \rangle}{6 \Delta t}.$$

The configurations obtained at the time point $t = 100 \, ns$ in the basic series of simulations (which correspond to the steady state of the system in the case that the temperatures are below the eutectic point) were used as the initial ones for calculating the diffusivities. Only the particles initially located at the distances no longer than 1 lattice period (approximately 4 angstrom) from the interface plane were taken into account to calculate the mean-square displacement (MSD) during the time span $\Delta t$. The latter was set short enough ($\simeq 300 ps$) to provide the linear dependence of MSD on time. It is evident, that due to anisotropy of the system and the dependence of the diffusion zone width on temperature, such a treatment will yield some averaged estimates for diffusivities in the boundary layer. Nevertheless, it provides a way to obtain important qualitative conclusions.

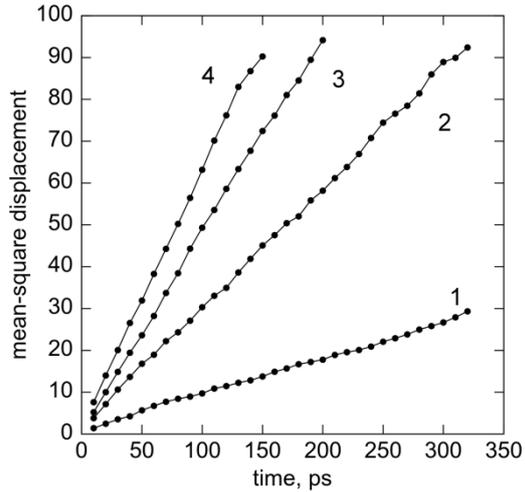

**Fig.4** Calculated MSD time dependence for copper atoms in boundary layer for the temperatures $T = 850(1); 900(2); 950(3); 1000(4)\ °K$

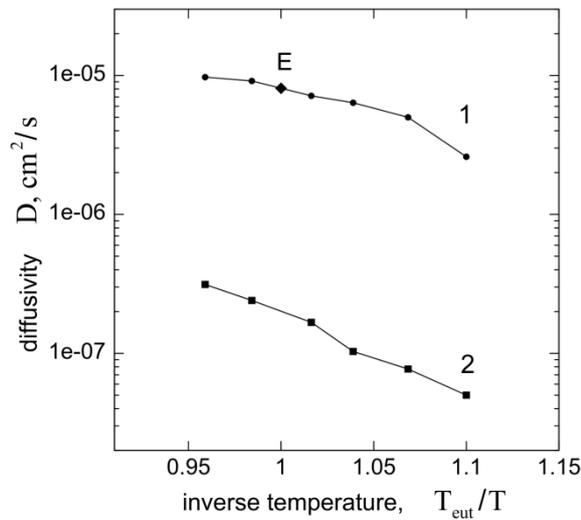

**Fig.5** Arrenius curves for the diffusion of copper atoms in the boundary zone (1) and in the crystal lattice (2). The eutectic point E is marked with a rhomb symbol. The mean-square error of the calculated numbers for the diffusion coefficients are within the range of 10%

The time dependences for MSD and the relevant Arrenius diagrams for copper atoms are given in Figs. 4 and 5. As follows from the results presented, the diffusivity of the copper atoms in the interface zone exceeds the numbers typical for crystal copper by two orders and increases as the temperature gets closer to the eutectic point. It should be noticed that the similar results can be obtained for silver atoms as well.

It is to emphasize that while passing the eutectic temperature (the point E in Fig. 5) the diffusion within the interface layer does not show any peculiarities. This means that during the temperature increase the complete CM is preceded by the similar process, which is, however, spatially localized in the vicinity of the interface. It is associated with the increase of the zone width, continuous growth of disorder and intensity of transport properties, up to the transition to the liquid phase.

**Conclusions**

The results obtained in simulations provide the basis for a number of important conclusions. MD computer simulations show that the contact of ideal crystal phases in binary AgCu eutectic is thermodynamically unstable. The equilibrium state in solid phase associated in eutectic with complete separation of the components is accompanied by the formation of steady disordered zone at the boundary between the crystal phases. Its width increases with rising the temperature, and reaches about 5 lattice periods near the eutectic point. The numbers for the diffusion coefficients of components within this interface zone turn out to be significantly higher than those observed within the crystal lattice and are close to the ones typical for liquid phase. Above the eutectic temperature, the steady interface zone does not form, instead, the complete CM in the system occurs. It should be pointed out that this phenomenon is quite analogous to the pre-melting phenomenon observed at grain boundaries in one-component systems (see, for instance, [10, 16, 17]), with the difference that in the case under consideration the CM in binary eutectic is examined.

The data obtained enable one to get a new insight into the results of simulations of contact phenomena based on the standard PFT model [11]. The equilibrium states of complete decomposition in binary eutectic with the liquid interface (three-phase states) observed in PFT can be regarded as an approximate or incomplete description of the above contact pre-melting phenomenon, which is localized, due to the limitations of the PFT model, within certain temperature range near the eutectic point. The disordered interface zone with high component diffusivities is described within this approach in terms of phase variable as a liquid state.

**Acknowledgements**

The authors acknowledge the support of European Office of Aerospace Research and Development, Project No. 118003 (STCU Project P-510).

**Conflict of interest**

The authors state that there is no any conflict of interest connected with the research presented.